# Challenges in designing edge-based middlewares for the Internet of Things: A survey

EDUARD GIBERT RENART, Rutgers University, USA
DANIEL BALOUEK-THOMERT, Rutgers University, USA
MANISH PARASHAR, Rutgers University, USA

The Internet of Things paradigm connects edge devices via the Internet enabling them to be seamlessly integrated with a wide variety of applications. In recent years, the number of connected devices has grown significantly, along with the volume and variety of data that is being generated by these devices at the edge of the network. An edge-based middleware is defined as a software that serves as an interface between the computational resources and the IoT devices, making communication possible among elements. Such middleware is required to provide the necessary functional components for sensor registration, discovery, workflow composition, and data pre-processing. In recent years, the landscape of the edge middleware platforms has grown exponentially, each of them with different platform requirements, architectures, and features. The core of this survey is a comprehensive review of existing edge middleware solutions. In this regard, we propose a four-layer architecture for the design of edge-based middleware, along with some design goals for each of the proposed layer. The paper concludes with some open challenges and possible future research directions.

## 1 INTRODUCTION

Cloud computing was introduced a decade ago with the promise of seemingly infinite computing resources available on demand [35]. This model has proved to be effective for scaling up search engines[36], social networks[93], and content service providers[30] to billions of users around the world. However, this centralized model is being challenged by the emergence of a new computing paradigm and associated technologies i.e. Internet of Things (IoT).

The Internet of Things paradigm (IoT) fosters the connection of large numbers of sensors to the network. According to Cisco systems [23], 500 billion IoT devices are expected to be connected to the Internet by 2030, and mearly 50% of the data produced worldwide will be generated by IoT sensors [73]. As the volume of data generated from the devices increases, moving data from the edge of the network to the Cloud might not be feasible due to bandwidth constraints [83]. Furthermore, as low latency and location-aware applications emerge [89], transfering all the data to the Cloud will not satisfy the low latency or location-aware constraints that the IoT applications expect. In addition, some applications, deal with sensitive and personal data, making it not possible to send the data to the Cloud due to privacy concerns [38]. For example, Toyota estimates that the amount of data flowing between vehicles and servers will reach 10 exabytes per month by 2025 [98]. Another example is commercial jets, which generate 10 TB of data for every 30 minutes of flight, making it impractical to transport all the data from the edge to the Cloud [19].

Edge computing has emerged as a potential approach for handling the large quantity of data generated by connected devices. It leverages the ability to execute computations and process data at the edge of the network, closer from the location of data producers. Edge computing leverages smaller servers or single board computers that are widely distributed close to the edge to improve delays. Edge middleware is the essential software stack that serves as an interface between the Cloud and the IoT devices, supporting data discovery, communication and processing between edge devices and cloud services.

Table 1. Edge Computing use cases, current limitations, and imperatives

| Applications | Example Use Case | Limitations | Requirements |
|---|---|---|---|
| Smart City [87] | Help autistic people navigate through large crowded spaces. | Hard to provide real-time directions as the analysis has to be performed in the Cloud. | Low Latency, Security, Geographically Distributed, Mobility, Scalability Reliability and Robustness |
| Disaster Recovery [85] | Need to timely determine building conditions after a natural disaster has struck. | Hard to perform real-time decision due to the need to send large volumes of data to the Cloud. | Low Latency, Geographically Distributed, Orchestration and Management |
| Scientific Observatories [103] | Large networked system of under water instruments to collect real-time data from the ocean. | Hard to deliver near real-time data to the end user due to the need to send large volumes of data to the Cloud. | Low Latency, Security, Geographically Distributed, Multi-Tenancy, Scalability |
| Video Analytics [32] | Video analytics for safety and security from public video cameras. | Hard to perform real-time analytics due to the need to send large volumes of data to the Cloud. | Low Latency, Security, Geographically Distributed, Scalability |

The realization of edge-based middleware platforms presents several conceptual and technical challenges. We believe that the seamless integration of edge and Cloud systems is one of the main challenges that prevent the efficient utilization of IoT. Without such an integration, developers must explicitly manage the platform as a unified set of resources to orchestrate computations, coordinate devices, and deliver data to users. As a result, there has been a substantial amount of research towards building edge-based middleware, addressing key crosscutting challenges, such as device discovery, scalability, and privacy and security. It is therefore important to understand the current state-of-the-art edge-based middleware and identify the gaps that may exist.

Multiple surveys on IoT middlewares have been published, such as [78, 84, 96]. To the best of our knowledge, existing surveys focus on specific classes of IoT systems and/or applications and as results only consider a subset of the currently available IoT middleware. Furthermore, none of these existing develop a reference architecture for edge middleware and use it to characterize and analyze the existing landscape. The contributions of this survey are as follows: First, we present a reference architecture for edge middleware. Second, we define a set of design goals for each of the layers of the reference edge middleware architecture and use it to characterize and compare and contrast state-of-the-art commercial and academic edge-based middleware solutions. Finally, based on this analysis, we review and present existing issues and gaps and highlight future research opportunities.

The rest of this survey is organized as follows. The motivating use cases for an integration of edge and Cloud resources are presented in Section 2. Section 3 presents a reference architecture for enabling end-to-end services in Edge Computing. We propose four architectural layers, along with individual design goals for each of the layers. Section 4 compares platforms and research projects according to the previously described reference architecture. Section 5 deals with challenges and future work for the realization of edge computing platforms. Section 6 summarizes and concludes this survey.

## 2 MOTIVATING USE CASES

IoT applications are present in several domains: Precision medicine, Urban mobility, and Healthcare. In this section, we highlight four different use cases described in both industry and academia that benefits from the IoT paradigm. Table 1 summarizes the scenario, limitations, and requirements of those use cases.

### 2.1 Smart City

The first use case is the smart cities for people with disabilities [87]. Large cities present navigation challenges for people with special needs such as visual impairment or Autism Spectrum Disorder (ASD). The main objective of this application is the use of IoT capabilities to provide location-aware services to improve travel experience and awareness of nearby events. In the smart city use case, the querying and coordination of multiple video cameras feeds from different locations incur high latencies, bandwidth congestion, and privacy concerns if dealt by sending the data to Cloud resources.

### 2.2 Disaster Recovery

Our second use case is associated to Disaster Recovery Manegement [85]. The disaster recovery use case consists of four phases: preparedness, response, mitigation and recovery. The recovery phase aims to restore the affected area to its previous state. This workflow focuses on the response phase by using a multi-stage generic response workflow that is executed between the edge and at the cloud. It starts by capturing LiDAR images of the affected zones followed by a pre-processing stage at the edge of the network, to determine if further post-processing is needed. If further processing is needed, data will be sent to the Cloud to perform a further post-processing. The workflow was extracted from the office of coastal management [1].

### 2.3 Scientific Observatories

Our third use case is focused on scientific observatories, in particular the Ocean Observatory Initiative (OOI) [103]. OOI is a networked ocean research observatory with arrays of sensors and autonomous underwater vehicles. This networked system of instruments provide scientists the means to collect data sets, and enables the examination of complex cyber-physical processes. The scientific observatory use case presents similar real-time constraints that prevents the sending large data products to the Cloud. This particularly affects timely delivery and transformation of data products into scientific insights.

### 2.4 Video Analytics

The last use case is the use of video analytics for safety and security [32]. A standard video camera produces between 553 Mbps and 1.24 Gbps for a minute of video recording. The ability to record 4K video on cameras will push that number to grow exponentially in the upcoming years. The traditional model to send all the data to the Cloud is not efficient enough to support such video data analytics [92]. Video Analytics pipelines need to be

Authors' addresses: Eduard Gibert Renart, Rutgers University, Piscataway, NJ, USA, egr33@rutgers.edu; Daniel Balouek-Thomert, Rutgers University, Piscataway, NJ, USA, daniel.balouek@rutgers.edu; Manish Parashar, Rutgers University, Piscataway, NJ, USA, parashar@ored.rutgers.edu.

performed using edge, in-transit and cloud resources in order to cater to low latency requirement for large-scale video streams [34].

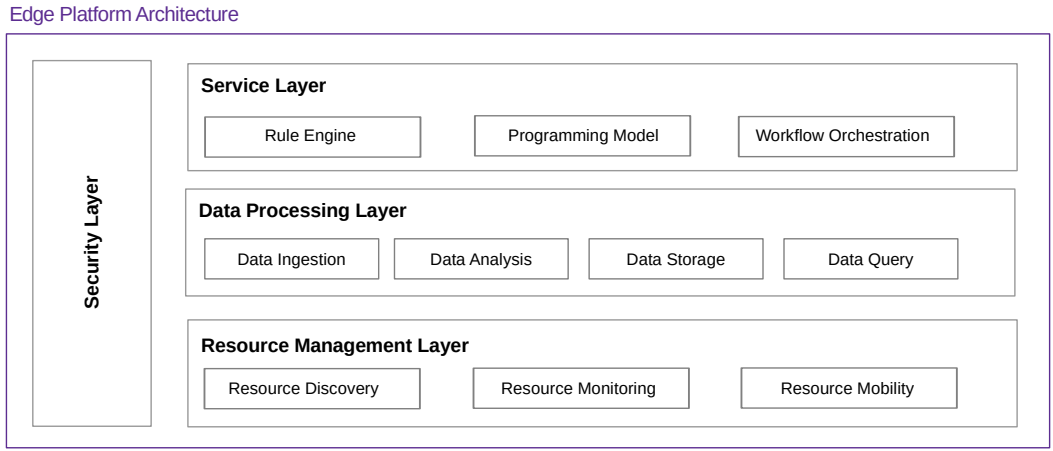

Fig. 1. Edge-based middleware reference architecture consisting of four layers, each of them with their respective components.

## 3 EDGE-BASED MIDDLEWARE ARCHITECTURE

Extensive research and development have been put into creating edge-based middleware systems. There are currently more than 100 edge-based middleware platforms in the market today and the number is continuously growing [82]. However, not every platform is designed with the same capabilities or architecture. Despite the diversity and large number of edge-based middleware systems, two common architectures emerge:

The majority of IoT platform's architecture follows the Cloud-centric approach. They are built on the premise that ingestion, management, and processing of IoT data can be done in the Cloud, without any edge computing capabilities. Some examples are: Particle Cloud [29], Salesforce IoT Cloud [2] and If This Then That [10].

The other approach is the end-to-end architecture or edge-based middleware architecture built on the premise that edge-processing can save huge costs to clients.

In this survey, we focus on the end-to-end or edge-based middleware architecture, since the Cloud-centric approach will not be able to satisfy the requirement of the IoT applications presented in section 2. In order to compare and contrast all the existing state-of-the-art edge-based middlewares, we carefully studied the requirements and limitations of the IoT applications and came up with a four-layer edge-based middleware framework that satisfies all the requirements and limitations of the IoT applications, and each of the middlewares should consist. Figure 1 presents the layers and components that need to be included in an edge-computing solution. The edge-based middleware architecture is composed of four separate layers: resource management, data processing, service , and security.

### 3.1 Resource Management Layer

The resource management layer is dedicated to the discovery, identification and allocation of available resources. The challenge of the resource management layer is in managing these limited and geo-distributed resources efficiently. The resource management layer consists of the following components: resource discovery, resource monitoring, and resource mobility. Table 2 summarizes the design goals of each of the works focused on the resource management layer.

Table 2. Design goals of the resource management layer components of the cited papers in this survey.

| Paper | Resource Discovery | | Resource Monitoring | | Resource Mobility | |
|---|---|---|---|---|---|---|
| | Distributed | Low Overhead | Distributed | Low Overhead | Distributed | Low Overhead |
| Paganelli et al. [79] | ✓ | ✓ | | | | |
| Liu et al. [68] | ✓ | | | | | |
| Cirani et al. [45] | ✓ | ✓ | | | | |
| Jara et al. [58] | | ✓ | | | | |
| Zhou et al. [105] | | ✓ | | | | |
| Tanganelli et al. [95] | | | ✓ | ✓ | | |
| Mäenpää et al. [70] | | | ✓ | ✓ | | |
| SEGUE [104] | | | | | ✓ | ✓ |
| Chaufournier et al. [42] | | | | | ✓ | ✓ |
| Farris et al. [51] | | | | | ✓ | ✓ |

*3.1.1 Design Goals.*
We suggest the following design goals to be considered when designing components in charge of resource management.

**Low Overhead:** The algorithms and protocols of the resource management layer need to offer low runtime overhead in performance-limited hardware platforms, in devices such as Rasperry Pi's, smartphones, or edge gateways.

**Distributed:** Ensuring a deployment and avalability of applications at scale with regards to the dispersed IoT and Edge devices implies a coordination of distributed components.

*3.1.2 Resource Discovery.* The resource discovery component is responsible for efficiently identifying and discovering the geo-distributed IoT sensors. The following are some of the work focused on the resource discovery component.

Paganelli et al. [79] present a service for discovering Internet of Things resources. The service uses a peer-to-peer approach along with distributed hash table (DHT) techniques to support the discovery of distributed resources, the system guarantees scalability, robustness, and maintainability. Paganelli et al. meet both design goals since they use a distributed architecture by the means of a P2P architecture to support the number of growing devices and offers a low-overhead algorithm.

Liu et al. [68] propose a distributed architecture for resource discovery, designed to be used in Machine-to-Machine applications. The architecture uses an overlay network composed of peer nodes to distribute workload,and eliminating the single point of failure. Liu et al. resource discovery mechanism only meet the distributed goal since the system is build using a peer-to-peer architecture to efficiently discover the resources in a decentralized manner. It does not meet the low overhead since it used HTTP to communicate and discover resources [77]. The reason being that HTTP runs on TCP, therefore it incurs all TCP connection overheads for connection establishment and closing [77].

Cirani et al. [45] also present a Peer-to-Peer architecture for service and resource discovery that can be applied for the Internet of Things applications. Cirani et al. resource discovery mechanism satisfies both goals since it used a distributed P2P architecture for discovering resources and it used CoAP a lightweight messaging system that uses UDP [77] for keeping track of all the resources.

Jara et al. [58] presents a centralized mechanism for discovering devices based on context and location. Jara et al. only satisfy the low overhead goal of the resource discovery mechanism since it uses a centralized architecture for discovering devices. It is well-known that centralized architectures have a single point-of-failure and present some scalability concerns when the number of IoT devices grows [68].

Zhou et al. [105] presents a service discovery algorithm and architecture designed for the Internet of Things. The work focuses on the context and location aware discovery. They first present an architecture called "Digcovery" to support the large number of IoT devices. And finally they present a search engine to offer query, look-up and filtering support. Zhou et al. only satisfy the low-overhead goal since the resource discovery mechanism claims that the algorithm has good scalability, and it can be applied to different fields. Only the domain ontology needs to be replaced.

*3.1.3  Resource Monitoring.* The resource monitoring component is responsible for controlling and managing hardware and software infrastructures. It also provides information and performance indicators for both platforms and applications to assist in the decision of allocating the resources. In addition, it monitors the state of the resources in the event of failure. The following are some of the work focused only on the resource monitoring component.

Tanganelli et al. [95] propose an edge-centric architecture that uses the CoRE Resource Directory interface and the CoAP protocol to enable resource monitoring and discovery for IoT applications. This approach is able to satisfy both design goals since it is distributed, in the means of a P2P network, and achieves low overhead since they run their experiments on emulated embedded devices and achieve millisecond latencies.

Mäenpää et al. [70] propose an architecture that focuses on the resource discovery and monitoring of wide area sensors and actuators. The architecture enables a federation of geographically distributed Wireless Sensor Networks (WSNs) using a peer-to-peer network. This approach satisfies both design goals.

*3.1.4  Resource Mobility.* The resource mobility component is responsible for moving computations between edge nodes in order to achieve the requirements of the IoT applications. The following are some of the work focused only on the resource mobility component.

SEGUE [104] is a migration system, that achieves optimal migration decisions by using the Markov Decision Process (MDP) to perform migration decisions. SEGUE meets all the design goals since it was carefully evaluated to showcase the real-time performance, scalability, and dynamicity by using real mobility trace of 320 taxis in Rome.

Chaufournier et al. [42] relies on multi-path TCP, an effort to use multiple paths to maximize resource usage and increase redundancy. This techniques aims at improving the migration time of virtual machines. Chaufournier et al. resource mobility approach also achieves all the goals since it proposed the uses of multi-path TCP and claims that increases the migration throughput by 6x and reduces the time by 50% in some cases.

Farris et al. [51],presents two Integer Linear Problem optimization schemes, with the pourpus of reducing the quality of service when performing migrations at the edge of the network. Farris et al. resource mobility also meets all the goals since it was designed to cope with the limitation of resource-constrained edge nodes and showcased the scalability in terms of users and the dynamicity of the algorithm.

## 3.2  Data Processing Layer

The data processing layer is in charge of the consolidation of data from multiple producers, along with its processing and delivery. Current approaches in data processing are known to be data-intensive process. The frequent operations on disk results in the inability to perform real-time data analytics when executed on edge constrained devices. The data processing layer consists of four components: ingestion, analysis, storage, and query. Table 3 summarizes the design goals of each of the works focused on the data processing layer.

Table 3. Design goals of the data processing layer components of the cited papers in this survey.

| Project | Data Ingestion | | | Data Analysis | | | Data Storage | | | Data Query | | |
|---|---|---|---|---|---|---|---|---|---|---|---|---|
| | Real-Time | Distributed | Scalable | Real-Time | Distributed | Scalable | Real-Time | Distributed | Scalable | Real-Time | Distributed | Scalable |
| Apache Kafka [63] | | ✓ | ✓ | | | | | | | | | |
| Mosquitto [46] | | ✓ | ✓ | | | | | | | | | |
| RabbitMQ [6] | | ✓ | | | | | | | | | | |
| ActiveMQ [7] | | ✓ | | | | | | | | | | |
| Heron [64] | | | | | ✓ | ✓ | | | | | | |
| Storm [97] | | | | | ✓ | ✓ | | | | | | |
| Flink [41] | | | | | ✓ | ✓ | | | | | | |
| MillWheel [31] | | | | | ✓ | ✓ | | | | | | |
| Spark [102] | | | | | ✓ | ✓ | | | | | | |
| ApacheEdgent [21] | | | | ✓ | ✓ | ✓ | | | | | | |
| LMC [80] | | | | ✓ | | ✓ | | | | | | |
| DataFlog [55] | | | | | | | ✓ | ✓ | ✓ | ✓ | ✓ | ✓ |
| FogStore [54, 72] | | | | | | | ✓ | ✓ | ✓ | ✓ | ✓ | ✓ |
| Moon et. al. [60] | | | | | | | | | | | ✓ | ✓ |
| IOTMDB [67] | | | | | | | | | | | ✓ | ✓ |

*3.2.1 Design Goals.*
We suggest the following design goals to be considered when designing components in charge of data processing.

**Distributed:** Similar to the resource component layer, the geographical dispersion of information processing features, computing capabilities, storage, and query requires a distributed design of components.

**Scalable:** The handle of 10,000 to 40,000 sensors transmitting at once consitutes the practical requirements of current implementations of Smart Cities (e.g, SmartSantander [4]). The ability to process a growing number of data streams is crucial to the system design.

**Real-Time:** As previously described in Section 2, achieving low latency responses as the number of messages increases needs to be realized on edge constrained devices or in the Cloud to ensure timely processing and delivery.

*3.2.2 Data Ingestion.*
The data ingestion component aggregates data from multiple producers and sources in order to enable processing through pipelines. The following works focused solely on the data ingestion component.

Apache Kafka [63] is one of the most popular frameworks available, it is an open-source framework used for building real-time data pipelines and streaming apps. Apache Kafka meets three of the four design goals, the reason Apache Kafka does not offer real-time processing at the edge of the network, because it was not designed to be deployed in constrained devices, as it was demonstrated in this work[86].

Mosquitto [46] is a lightweight open-source publish/subscribe messaging broker designed for the Internet of Things. It implements the lightweight MQTT protocol, to transport the messages, making it suitable for low-power devices. Even though Mosquitto was created for the need to achieve real-time message handling, Scalagent published a survey where they stress test the Mosquitto and show that it can succeed at handling 60,000 publishers but it requires high transmission latency and high CPU usage [90]. For those reasons Mosquitto only satisfies the scalability and distributed and scalable design goals.

RabbitMQ [6] is also a lightweight publish/subscribe messaging broker, designed to be deployed in the cloud. Similarly to Mosquitto, RabbitMQ in the scaleagent tests shows that it can only handle 8,000 publishers producing 8,000 messages per second, and is not able to achieve real-time analytics [90]. In this case RabbitMQ only satisfies the distributed goal since it can only support 8,000 publishers, and as mentioned earlier, current city-scale experimental research facilities envision the deployment of 20,000 to 40,000 sensors [4].

ActiveMQ [7] its an open-source messaging broker, that supports numerous industry-standard protocols. ActiveMQ also suffers from the same problems as RabbitMQ since it has high message transmission latency and cannot handle more than 20.000 publishers [90].

### 3.2.3 Data Analysis.

Data analysis is the process of analyzing large volumes of data to discover useful information and perform informed decisions. The following are some of the work focused on the data analysis component.

Heron [64] is a real-time analytics platform developed by Twitter. It is designed for speedy performance, low latency, isolation, and reliability. Heron meets all the design goals except for the real-time data analytics at the edge because it was designed to be deployed in large clusters at the core of the network.

Apache Storm [97] is an open-source distributed real-time stream processing system. Similarly, Storm was also designed to be deployed in the Cloud.

Flink [41] is a distributed processing engine for performing stream processing applications over unbounded and bounded data streams. Flink was also designed to be deployed in the Cloud and not for the edge.

MillWheel [31] is a framework for building low-latency data-processing applications that was designed and build by Google. MillWheel, just like Heron, Storm, and Flink, was designed to be deployed in large clusters in the Cloud.

Spark [102] is an open-source distributed general-purpose stream processing and batch processing framework witch allow to perform in-memory analytics. Spark, just like Heron, Storm, and Flink, was designed to be deployed in large clusters in the Cloud.

Apache Edgent [21] is a micro-kernel framework designed to be deployed in small footprint edge devices, enabling local, real-time analytics at the edge of the network. Apache Edgent is a stream processing engine that was designed to be deployed on edge devices, allowing it to achieve all the design goals.

LMC [80] enables cross-platform code execution on constrained IoT devices. LCM meets the real-time and the scalable design goals since it was designed to constrained devices , but it doesn't meet the distributed goal since there is no currently not supported.

### 3.2.4 Data Storage.

Due to the ever-increasing deployment of bandwidth-intensive IoT platforms (especially cameras), there is an increasing pressure on the bandwidth to transport data back and forth between the edge and the Cloud. There is a need for a more efficient management and computation of the data at the edge of the network. Building a storage system on an edge computing infrastructure has its own set of particular challenges. The wide geo-distribution and heterogeneous and constrained natures of this infrastructure require data-partitioning and replication policies that are commensurate with the latency requirements of the applications. The following are some of the work focused on the data storage component.

DataFlog [55] is a distributed indexing mechanism that performs data placement (both among edge nodes, and between the edge and the Cloud) based on spatiotemporal attributes to support efficient queries involving multiple edge nodes. DataFlog is able to achieve all the design goals for the data storage layer since it uses distributed indexing mechanism, it supports efficient queries and it can scale since it uses a P2P network and can be deployed in any environment.

FogStore [54] [72] is distributed key-value storage system tailored for the edge of the network. FogStore uses a fog-aware replica placement, and a context-sensitive differential consistency strategies to satisfy the requirements of the Edge and the Fog. FogStore was designed by the same authors of DataFog and it also meets all the design goals, just like DataFlog.

### 3.2.5 Data Query.
Similarly to the data storage, once data has been stored it needs to be accessed as well. The following are some of the research work on creating edge query systems. The following are some of the work focused on the data query component.

Moon et al. [60] propose a data management and searching system based on blockchain which ensures security. Moon et al. are able to meet all the goals for the data query layer except for the real-time design goal since they are using the blockchain Proof-of-Work consensus algorithm and it is known that the time to perform a computation does not increase linearly as the number of nodes increases.

IOTMDB [67] is an IoT storage solution based on NoSQL (Not Only SQL), to solve the storage and management problems of large volumes of IoT data. IOTMDB is able to satisfy all the design goals except for the real-time, since storing 1,000 records can take up to 2 seconds since other frameworks such as RocksDB can store 1,000 records in less than 60 ms [86].

Table 4. Design goals of the service layer components of the cited papers in this survey.

| Paper | Rule Engine | | Programming Model | | Workflow Orchestrator | | |
|---|---|---|---|---|---|---|---|
| | Scalable | Low Overhead | Expressive | Extensible | Dynamic | Scalable | Low Overhead |
| Chui et. al. [65] | ✓ | ✓ | | | | | |
| Lica et. al. [71] | ✓ | | | | | | |
| Mobile-Fog [57] | | | ✓ | ✓ | | | |
| Rabel [88] | | | ✓ | ✓ | | | |
| Fabryq [50] | | | ✓ | ✓ | | | |
| FogFlow [44] | | | ✓ | ✓ | | | |
| Eidenbenz et al. [47] | | | | | | | ✓ |
| Taneja et al. [94] | | | | | | ✓ | ✓ |
| DROPLET [49] | | | | | ✓ | ✓ | ✓ |
| Ghosh et al. [52] | | | | | ✓ | | |

## 3.3 Service Layer
The service Layer defines an application's set of available operations to the end user. The service layer is composed of three components: rule engine, programming model, and workflow orchestrator. Table 4 summarizes design goals of each of the works focused on the service layer.

### 3.3.1 Design Goals.

We suggest the following design goals to be considered when designing components in charge of data processing.

**Low Overhead** Components are required to provide timely analysis and data queries when deployed in performance-limited hardware platforms.

**Scalable** The service layer components need to offer good scalability, since there is going to be a large number of rules and a large number of operators that need to be placed.

**Dynamic** This design goal only applies to the workflow orchestrator. The workflow ochestrator needs to be able to orchestrate the workflows based on the runtime characteristics of the nodes.

**Expressive** This design goal only applies to the programming model. The programming model needs to be easy to express ideas, algorithms, and tasks in an easy-to-read and succinct way.

**Extensible** This design goal also only applies to the programming model. The programming model needs to flexible enough that if new capabilities are needed, they can be added to the software without major changes to the underlying architecture.

### 3.3.2 Rule Engine.

The Rule Engine makes it possible to evaluate data, perform decisions and trigger actions. The following are some of the work focused on the rule engine component.

Chui et al. [65] propose a rule-based system designed to support heterogeneous IoT devices. The rule-based system is based on Event-Condition-Action (ECA) rule mechanism with SOAP technology. Chui et al. rule engine satisfies all the design goals since they use an Event-Condition-Action (ECA) pattern which allows them to scale the system as the number of rules grows and achieve low overhead.

Lica et. al [71] propose a rule-based architecture that addresses the main issues involved in application management in the Internet of Things. Lica et al. approach satisfies the expressive and extensible goals since further developments are necessary to improve the architecture effectiveness before its final implementation is carried out.

### 3.3.3 Programming Models.

Due to the high dynamicity of edge resource, heterogeneity of Cloud and edge resources deploying low latency and scalable applications can be tricky. For this reason, there is a need for high-level programming models that simplify the development of IoT applications across the edge and the Cloud. The following are some of the work focused on the programming model component.

Mobile-Fog [57] is a high-level programming model designed for applications that require large number of sensors and actuators and they are latency-sensitive. Mobile-Fog only satisfies both design goals since it uses a high-level API to program the sensors, making it easy to learn.

Ravel [88] proposes a programming model to program applications across embedded devices, edge nodes and cloud nodes by using an extension of the Model-View-Controller architecture. Ravel satisfies both design goals since it uses a high-level API.

Fabryq et al. [50] propose a proxy programming model to find and control sensors and actuators. Fabryq et al. approach also satisfies both design goals since it uses Javascript as the main programming language and also has a high-level API to program the sensors, making it easy to learn.

FogFlow [44] is a programming model that extends the dataflow programming model, allowing developers fast and easy development of edge and fog applications. FogFlow satisfies both design goals since it extends the Cloud dataflow programming model and makes it suitable for the edge environment, making it easy to learn.

### 3.3.4 Workflow Orchestrator.

The Workflow Orchestrator consists of defining how to accommodate the application components (i.e., operators) on the available resources of the network topology to optimize one or more performance metrics [53]. The main challenge is to decide how to split the operators between the edge and Cloud in order to minimize the overall completion time. The workflow placement has been proved to be at least NP-Hard [40]. The following are some of the work focused only on the workflow orchestrator component.

Eidenbenz et al. [47] present an algorithm for the Series-Parallel-Decomposable Graphs (SPDG). Eidenbenz et al. only satisfy the real-time design goals since its only a theoretical approach.

Taneja et al. [94] propose an approach for deploying application across Cloud and edge resources by using a Module Mapping Algorithm. Taneja et al. meet all the design goals except for the dynamicity since the approach doesn't take into consideration network connectivity or failure of nodes.

DROPLET [49] is an algorithm, that partitions tasks across the edge and Cloud resources, while minimizing the total completion time. DROPLET achieves all the design goals since it is able to react and adapt to dynamic network events and is capable of performing real-time decisions and scale polynomially with increasing the number of operators to place.

Ghosh et al. [52] propose a Genetic Algorithm (GA) meta-heuristic for distributing analytics across edge and Cloud resources to support IoT applications. The main goal of the genetic algorithm is to minimize the end-to-end latency. Ghosh et al. only meet one of the design goals since it takes between 1 - 26 seconds for placing 1 - 50 operators, making it not real-time or scalable when the number of operators grows.

Table 5. Design goals of the security layer components of the cited papers in this survey.

| Paper | End-to-End Security | Data Privacy |
|---|---|---|
| Lu. et al. [69] | | ✓ |
| Shi et al. [91] | | ✓ |
| Behrens et al. [39] | ✓ | |
| Mukherjee et al. [76] | ✓ | |
| Kothmayr et al. [62] | ✓ | |

## 3.4 Security Layer

The fourth and last layer is the Security Layer, which consists of keeping the data generated by thousands of IoT devices private and secure. The following are some of the work focused on the end-to-end security component. Table ?? summarizes the work focused on the security layer.

### 3.4.1 Data privacy.

Since the IoT produces large volumes of data easily available privacy protection in IoT its a challenge. The following are some of the work focused only on the data privacy component.

Lu et al. [69] present a lightweight privacy-preserving data aggregation scheme designed to be used in constrained devices. The proposed aggregation schema uses the homomorphic Paillier encryption, Chinese Remainder Theorem, and one-way hash chain techniques to aggregate data.

Shi et al. [91] propose an algorithm that allows users to upload encrypted data to an untrusted aggregator, and allows the aggregator to decrypt statistics for each time interval.

### 3.4.2 End-to-End Security.

To preserve user's privacy it is critical that the communication links between the IoT devices and the servers are secured, in order not to leak personal data. The following are some of the work focused on the end-to-end security component.

Behrens et al. [39] present an end-to-end Application Security Layer for a fast and secure communication, designed for the Internet of Things devices. The Application Security Layer makes use of TLS over JSON, providing a way for end-to-end communication.

Mukherjee et al. [76] present the design a middleware specifically designed for securing the Internet of Things, featuring a flexible security configuration that allows developers to tailor the security to the needs of the application.

Kothmayr et al. [62] present a two-way authentication security scheme for the Internet of Things (IoT) applications. The proposed scheme relies on the public key cryptography (RSA).

## 4 COMMERCIAL AND ACADEMIC EDGE MIDDLEWARE SYSTEMS

In this section, we analyze the similarities and differences amongst all the currently available edge-based middleware systems that implement one or more of the layers of our edge-based middleware architecture. To do so we use the proposed edge platform architecture and the goals of each of the layers described in the previous section. Tables 6,7,8,9 summarize and offer more details on all the edge middleware surveyed systems, including the design goals that each component satisfies.

AWS Greengrass [33] is a software stack that allows to locally run computations, messaging, data caching, sync, and Machine Learning capabilities on devices in a secure way. AWS Greengrass consists of all the four layers presented in section 3. The main limitations of AWS Greengrass are the centralized architecture of the resource management layer, the lack of storage and query of the data processing layer, the use of a similar MQTT broker to Mosquitto for data ingestion violating the real-time design goal for the data processing layer, and the lack the workflow orchestration component in the service layer, leaving it to the end-user for the management and provisioning of the workflows.

Azure IoT Edge [75] is a collection of services designed to create end-to-end IoT applications on Azure Cloud. This service is meant for analyzing data at the edge of the network, instead of in the Cloud. Azure IoT, similarly to AWS, takes security very seriously, and uses certificate-based authentication as the primary mechanism for authentication for the Azure IoT Edge platform. Azure IoT also implements all four layers proposed in section 1. Azure IoT only misses two components: the first one is the workflow orchestrator from the service layer and the second one is the data privacy at the security layer. Azure IoT uses a similar MQTT broker to Mosquitto making it not able to achieve real-time analytics.

EAaaS [99] is an analytics service that enables real-time edge analytics in IoT scenarios. The main focus of the EAaaS is the uses of a unified rule-based analytic model to simplify the user's programming efforts. In addition, they put a great amount of attention on making the system as lightweight and scalable as possible. EAaaS implements two of the four layers; it does not implement the resource management layer or the security layer and misses some components on the layers that it implements. The first components missing are from the data processing layer: EAaaS does not allow the storage or query of data at the edge of the network. From the service layer, EAaS does not implement the workflow orchestrator, forcing the end-user to decide where to place computations to achieve optimal performance.

Google Cloud IoT Edge [20] is a collection of services that allows users to manage, and consume IoT data from distributed devices at a large scale, and take actions as needed. Google Cloud IoT Edge follows the same path as the AWS Greengrass, implementing all four layers but missing some critical components on some of the layers. Google Cloud IoT Edge does not support the ability to store or query at the edge of the network. In addition, just like all the commercial systems surveyed so far, it also implements a similar broker to Mosquitto, violating the real-time design goal. Google Cloud IoT Edge does not offer the ability to orchestrate application between the edge and the Cloud.

Everyware IoT [14] is a comperical platform that provides an end-to-end IoT platform with proprietory software and hardware solutions. Everyware IoT implements all four layers but misses some critical components in all the layers. Similar to AWS, Azure, and Google, it lacks the query and storage support at the edge of the network and uses a similar Mosquitto broker for the data ingestion. Additionally, Everyware IoT lacks the rule engine of the service layer making it not possible to trigger or react to events that happen at the edge of the network.

Predix [48] is General Electric's commercial software platform for the collection and analysis of data from industrial machines. Predix implements all four layers but misses some critical components. Predix does not offer resource monitoring, storage, or query. In addition, it also doesn't offer the ability to orchestrate workflows between the edge and the cloud.

Bosch IoT [22] is a commercial end-to-end IoT platform that consists of multiple Cloud-enabled services and software packages. Bosch IoT implements all four layers but misses some components. Bosch IoT does not offer the rule engine or the workflow orchestrator of the service layer, and just like all other commercial systems it also implements a similar broker to Mosquitto.

Yanzi [5] is a commercial IoT platform designed to optimize office costs and productivity. Yanzi implements three of the four layers, lacking the resource management layer and some critical components on other layers. In the service layer, Yanzi misses the rule engine and the workflow orchestrator, and in the security layer, it misses the data privacy component.

R-Pulsar [85, 87] is an academic platform software stack that lets you run local analytics, messaging, data storage, and data querying capabilities on edge devices. R-Pulsar is the only one that satisfies all four layers with the most design goals. In addition, is the only software stack that has a full memory-mapped pipeline making it truly real-time. Also, it's one of the few that offers a unified architecture between the edge and the core, allowing it to seamlessly program the edge and the core. A limitation that the majority of the software stacks present is a split platform architecture between the edge and the core, leaving the end user to manage the scalability, replication, and distribution to the end user. For platforms that use a single architecture such as R-Pulsar, the system takes care of it so the user can focus on developing the application. R-Pulsar is also the only one to offer any application objectives, all the other software do not any application objectives.

FogHorn [15] is a commercial software platform that enables to run advanced analytics and machine learning applications at the edge of the network. FogHorn implements all four layers but misses some critical components in some of the layers. In the data processing layer it misses the data storage and query components, not allowing the storage or query of data at the edge of the network. In the service layer, it misses the workflow orchestrator making the end user responsible for the management and provisioning of the resources and workflows. In the security layer, it misses the data privacy component.

GeeLytics [43] is an academic platform, which can perform real-time analytics either at the edge edge, or in the Cloud in a dynamic manner. Geelytics was designed to emphasize the service layer, in particular, the workflow orchestration component. Geelytics enables developers to run stream processing applications across the edge and the Cloud, without the need to consider where each task is located. GeeLytics implements two of the four layers, not implementing the security and the resource management layer. In addition, GeeLytics lacks the rule engine in the service layer and makes use of Mosquitto or Apache Kafka as the data ingestion data processing layer making it hard to scale or perform real-time analytics at the edge of the network.

Fogflow [44] is the evolution of GeeLytics, an academic framework that orchestrates workflows over the Cloud and the edge based on various context, including system context. For this second iteration they improved their workflow orchestration mechanism, added the missing rule engine component, and implemented the resource management layer. Some of the drawbacks existing on the previous version still have not been addressed, such as the use of Mosquitto or Kafka as the data ingestion component, limiting the scalability and the performance at the edge of the network.

OpenMTC [13] is a commercial open-source implementation of an IoT/M2M middleware with the focus on providing a standard-compliant platform. OpenMTC implements three of the four layers, missing the resource management layer. In the data processing layer it does not allow the storage or query of data at the edge of the network, and just like any other commercial approach, it uses a similar MQTT broker for the data ingestion. In addition in the service layer, it misses the workflow orchestration.

SiteWhere [8] is an industrial open-source platform, that uses a multi-tenant microservice-based infrastructure. SiteWhere implements all four layers and only misses very few components on some of the layers. In the service layer, it lacks the workflow orchestration and in the security layer, it lacks data privacy.

SmartThings [11] is a commercial IoT platform designed for the smart houses. SmartThings implements three of the four layers missing the resource management layer and lacks some major components in some layers. In the data processing layer lacks the ability to store or query data at the edge of the network. In the service layer, it also lacks the workflow orchestration.

Kaa [16] is a commercial-grade IoT platform that is fully customizable. Kaa is one of the commercial systems more complete, implementing all four layers and missing very few components in some layers. The main drawback of Kaa is the lack of workflow orchestration between the edge and the Cloud, and the lack of data privacy in the security layer.

Samsung Artik [26] is a commercial IoT platform that focuses on unifying hardware, software, the cloud and the edge as a single ecosystem. Samsung Artick implements three of the four layers, missing the resource management layer. The main drawback is the lack of two of the key components in the data processing layer: the storage and query components. In addition, like all other commercial systems, Artick uses an MQTT broker similar to Mosquitto for the data ingestion component.

Ayla Network [9] is a commercial end-to-end IoT platform that includes a completely managed Cloud service. Ayla implements all the layers except for the resource management layer. In the data processing layer, it lacks the data storage and query and it uses an MQTT broker for the data ingestion layer. In addition, it also lacks the workflow orchestration component.

Altair SmartWorks [18] is a commercial platform designe as a Platform as a Service (PaaS) for Internet of Things projects, to collect data from objects, store it and build applications. Altair SmartWorks consists of three of the four layers, missing the data management layer. In the data processing layer, it lacks the ability to store and query data at the edge of the network. In also does not offer the ability to orchestrate workflows between the edge and the Cloud.

EdgeX [28] is a commercial open-source IoT microservice framework that allows end uses to chose their sensors from a large ecosystem of 3rd party offerings. EdgeX implements all four layers but lacks some of the components in most layers. In the data processing layer, EdgeX does not support the data storage or query. In addition like all other commercial systems, EdgeX uses an MQTT broker for the data ingestion violating the real-time design goal. In the service layer, it lacks the workflow orchestration.

PiCasso [66] is an academic orchestration engine that deploys services based on specifications and resources availability. PiCasso implements all the layers except for the security layer. PiCasso puts a lot of emphasis in the service layer more, in particular, the workflow orchestration component. PiCasso lacks the storage and query components of the data processing layers.

Hua-Jun Hong et al. [56] is an academic fog computing platform that that focuses on the task distribution between the edge and the cloud. Hua-Jun Hong et al. approach implements three of the four layers, missing the security layer. In addition, it misses most of the components in all layers, since the main focus of this platform is to make deployment decisions to maximize the number of satisfied IoT analytics (operator deployment problem). In the data processing layer lacks the ability to store and query data at the edge of the network.

Cloud4IoT [81] is an academic platform that focuses on automatically deploying and orchestrating IoT applications. Cloud4IoT implements all the layers except for the security layer. Cloud4IoT to ease the code interoperability between the edge and the Cloud, to do that relies on commercial software that was designed to be deployed on a large cluster, making it hard to achieve real-time analytics at the edge of the network.

Nebulae [25] is a commercial end-to-end IoT platform, which the main focus in interoperability and interportability. Nebulae implements three of the four layers missing the resource management layer. In the data processing layer, it lacks the ability to store and query data at the edge of the network. In the service layer, it lacks the rule engine and the workflow orchestrator.

FogGIS [37] is an academic framework for improving throughput and reducing latency for analysis of geospatial data. FogGIS implements all the layers except for the resource management layer. In addition, FogGIS data processing layer relies on a commercial system designed to be deployed on the Cloud not at the edge with constrained devices, making it hard to achieve real-time analytics.

FOG-engine [74] is an academic end-to-end platform for processing real-time analytics of data near where it is generated. FOG-engine implements two layers, not implementing the resource management and security layers, missing some key components on most layers. In the service layer, it lacks the orchestration and management of resources and workflows.

CEFIoT [59] is an academic end-to-end fault-tolerant architecture that reuses Cloud technologies at the edge of the network. CEFIoT implements two of the four layers, missing the resource management and security layers. In the service layer, it lacks the rule engine.

SAVI-IoT [61] is an academic self-managing programmable IoT platform that leverages both Hybrid Virtual Machines (HVV) and container isolation techniques to manage IoT applications. SAVI-IoT, just like CEFIoT, misses the same layers. The main difference is that SAVI-IoT does not offer a rule engine or a workflow orchestration. Another drawbacks of SAVI-IoT uses Kafka as the data ingestion component and Spark for the data analyses layer making them violate the real-time analytics at the edge of the network when deployed on constrained devices.

Foggy [101] is an academic architectural framework and software platform based on open-source technologies. Foggy main focus is the orchestration of application across the edge and the cloud. Foggy implements three of the four layers, missing the resource management layer. One of the main drawbacks of Foggy is the use of containers for orchestrating resources between the federated resources, making it no able to perform real-time analytics at the edge of the network. In addition, it lacks the ability to support storage and query at the edge of the network.

ISYMPHONY [100] is an academic orchestration framework designed for scaling real-time and on-demand IoT services. ISYMPHONY implements three of the four layers missing the security layer. ISYMPHONY focuses on the service layer in particular in the workflow orchestration layer. In the data processing layer, it lacks the data storage and query components. In addition, it lacks the rule engine in the service layer.

Macchina.io [17] is a commercial IoT SDK that allows to connect sensors, actuators, Cloud services, mobile devices, and humans. Macchina.io implements three of the four layers, missing the resource management layer. In the service layer, it doesn't offer a rule-based engine or the workflow orchestration. In addition, Macchina.io relies on an MQTT broker similar to Mosquitto for the data ingestion layer.

Clearblade [3] is a commercial IoT platform to build scalable, secure enterprise IoT solutions. Clearblade implements three of the four layers, missing the resource management layer, and just like every other system it implements Mosquitto as their data ingestion component.

IBM Watson IoT Platform [24] is a commercial IoT platform that can connect and control IoT sensors, appliances, homes, and industries. The IBM Watson IoT Platform relies on the cloud to distribute and manage the edge analytics. IBM Watson IoT Platform implements all four layers proposed but misses the data storage and data query components of the data processing layer. Just like every other commercial systems surveyed above, it uses Mosquitto as their data ingestion component making it not scalable and real-time.

## 5 CHALLENGES AND FUTURE WORK

Although there are numerous edge-based middleware solutions currently available, several open challenges remain. From the research papers surveyed above, we have identified the following research gaps that need to be addressed in order to advance in the field.

**Energy Management:** A study published in 2017 determined that due to the large number of IoT devices connected to the internet by 2025 they will consume 20% of all the worldwide electricity consuption [27]. For those reasons there is a need to implement energy management policies.None of the surveyed systems have the ability to quantify the amount of energy spent or the ability to schedule computations in an energy-efficient manner. Energy management needs to be incorporated in the service layer in order to be able to schedule computations based on the energy consumption. A large amount of research exists focused on modeling and optimizing the energy consumption in the Cloud, but there is limited research targeting edge computing. A possible research direction and a potential starting point is to design tools that allow developers to reason about energy consumption, or implement a workflow orchestrate that schedules computations between the edge and the Cloud while considering energy consumption of the computation and the communication.

**Real-Time:** Real-time a very important aspect of IoT data since data needs to processed in the right time frame in order to extract the value of the data. In most of the middleware systems surveyed in this paper, the data processing layer components rely on software stacks designed to be run on large server-based platforms instead of building on software stacks designed for constrained devices, limiting the ability to support real-time analytics at the edge of the network. For example, a large number of edge-based data processing solutions are based on Cloud-based software stacks such as Apache Kafka and Apache Storm. As a result, edge-based real-time data processing requires novel lightweight data ingestion components as well as data analytics frameworks that are specifically designed for edge devices.

**Mobility:** Few of the surveyed systems offer the ability to monitor sensor location and move the computations closer to it. Mobility needs to be incorporated at the service layer in order to move computations so they can achieve the low latency requirements that IoT applications need. There is an extensive amount of existing research focused on live migrations of virtual machines in the Cloud, but there is a lack of research exploring live migrations in an edge and Cloud environment. A potential research direction is the development of live migration techniques that are anticipatory and are based on forecasting movement in the near future, reducing down time due to migration in heterogeneous environments.

**Security:** IoT data differentiates itself from any other type of data due that is mostly built upon personal and highly sensitive data. For those reasons security is an important research topic. Even though a large number of the systems surveyed already offer some flavor of security, the majority of the protocols were designed to run on high-end server-based clusters where there is no constraint on how many processing cycles they can use. There is a need for algorithms that provide strong security guarantees, while still being suitable for constrained environments. A possible research direction is developing algorithms that provide sufficient security guarantees to handle sensitive IoT data while consuming acceptable levels of CPU cycles and energy.

## 6 CONCLUSION

In this paper, we analyzed the current requirements, limitations, and state-of-the-art edge-based middlewares. To do that we presented four IoT applications described in both industry and academia that benefit from the IoT paradigm. From the use cases we extracted the current requirements and limitations of the IoT applications, and we designed and proposed a four-layer edge-based middleware that satisfied all the needs. The proposed four-layer edge-based middleware consists of the resource management layer, the data processing layer, the service layer, and finally the security layer. In addition, each of the layers contains three or more components, each of which are required for satisfying the functionality of each layer. Then the proposed four-layer edge-based middleware was used to compare and contrast the academic and industrial sate of the art of the edge-based middleware systems. Finally, after having surveyed and carefully analyzed the current state-of-the-art edge-based middlewares, we outlined some of the existing challenges and future research work that need to be performed to drive innovation within the edge-middleware domain.

## ACKNOWLEDGMENTS

This research is supported in part by NSF via grants numbers OAC 1339036, OAC 1441376, OCE 1745246, OAC 1826997, OAC 1835692, and OAC 1640834, and was conducted as part of the Rutgers Discovery Informatics Institute (RDI[2]).

Table 6. (Appendix 1) Four Layer and Components for all the commercial and academic edge middleware available.

| System | Data Processing Layer | | | | Resource Management Layer | | | Service Layer | | | Security Layer | |
|---|---|---|---|---|---|---|---|---|---|---|---|---|
| | Data Ingestion | Data Analysis | Storage | Data Query | Resource Discovery | Resource Monitoring | Resource Mobility | Workflow Orchestrator | Rule Engine | Programming Model | Data privacy | End-to-End Security |
| AWS Greengrass [33] | ✓ | ✓ | | | | ✓ | | | ✓ | ✓ | | ✓ |
| Azure IoT Edge [75] | ✓ | ✓ | ✓ | ✓ | | ✓ | | | ✓ | ✓ | | ✓ |
| EAaaS [99] | ✓ | ✓ | | | | | | | ✓ | ✓ | | |
| Google Cloud IoT [20] | ✓ | ✓ | | | ✓ | ✓ | | | ✓ | ✓ | | ✓ |
| Everyware IoT [14] | ✓ | ✓ | | | ✓ | ✓ | | | | ✓ | | ✓ |
| Predix [48] | ✓ | ✓ | | | ✓ | ✓ | | | ✓ | ✓ | | ✓ |
| Bosch IoT [22] | ✓ | ✓ | ✓ | | ✓ | ✓ | | | | ✓ | | ✓ |
| Yanzi et. al. [5] | ✓ | ✓ | | | | | | | | ✓ | | ✓ |
| R-Pulsar [85, 87] | ✓ | ✓ | ✓ | ✓ | ✓ | ✓ | ✓ | ✓ | ✓ | ✓ | | ✓ |
| FogHorn [15] | ✓ | ✓ | | | | ✓ | | | ✓ | ✓ | | ✓ |
| GeeLytics [43] | ✓ | ✓ | | | | | | | | ✓ | | |
| Fogflow [44] | ✓ | ✓ | | | ✓ | | | ✓ | ✓ | ✓ | | ✓ |
| OpenMTC [13] | ✓ | ✓ | | | | | | | ✓ | ✓ | | ✓ |
| SiteWhere [8] | ✓ | ✓ | ✓ | ✓ | ✓ | ✓ | | | ✓ | ✓ | | ✓ |
| SmartThings [11] | ✓ | ✓ | | | | | | | ✓ | ✓ | | ✓ |
| Kaa [16] | ✓ | ✓ | ✓ | ✓ | ✓ | ✓ | ✓ | | ✓ | ✓ | | ✓ |
| Samsung Artik [26] | ✓ | ✓ | | | | | | | ✓ | ✓ | | ✓ |
| Ayla Network [9] | ✓ | ✓ | | | | | | | ✓ | ✓ | | ✓ |
| Altair SmartWorks [18] | ✓ | ✓ | | | | | | | ✓ | ✓ | | ✓ |
| EdgeX [28] | ✓ | ✓ | | | ✓ | ✓ | | | ✓ | ✓ | | |
| PiCasso [66] | ✓ | ✓ | | | ✓ | ✓ | | ✓ | | ✓ | | |
| Hua-Jun Hong et. al. [56] | ✓ | ✓ | ✓ | ✓ | ✓ | ✓ | | | | ✓ | | |
| Cloud4IoT [81] | ✓ | ✓ | | | ✓ | | | ✓ | | ✓ | | |
| Nebulae [25] | ✓ | ✓ | | | | | | | | ✓ | | ✓ |
| FogGIS [37] | ✓ | ✓ | ✓ | ✓ | | | | | | ✓ | | |
| FOG-engine [74] | ✓ | ✓ | | ✓ | | | | ✓ | | ✓ | | |
| CEFIoT [59] | ✓ | ✓ | ✓ | ✓ | | | | ✓ | | ✓ | | |
| SAVI-IoT [61] | ✓ | ✓ | ✓ | ✓ | | | | | | ✓ | | |
| Foggy [101] | ✓ | ✓ | | | | | | ✓ | ✓ | ✓ | ✓ | ✓ |
| ISYMPHONY [100] | ✓ | ✓ | | | | | | ✓ | | ✓ | | |
| Macchina.io [17] | ✓ | ✓ | ✓ | ✓ | | | | | | ✓ | | ✓ |
| Clearblade [3] | ✓ | ✓ | ✓ | ✓ | | | | | ✓ | ✓ | | ✓ |
| IBM Watson IoT [24] | ✓ | ✓ | | | ✓ | ✓ | | | ✓ | ✓ | | ✓ |

Table 7. (Appendix 1) Resource management layer and design goals for all the commercial and academic edge middleware available.

| System | Resource Discovery | | Resource Monitoring | | Resource Mobility | |
|---|---|---|---|---|---|---|
| | Distributed | Low Overhead | Distributed | Low Overhead | Distributed | Low Overhead |
| AWS Greengrass [33] | | ✓ | | ✓ | | |
| Azure IoT Edge [75] | | ✓ | | ✓ | | |
| Google Cloud IoT [20] | | ✓ | | ✓ | | |
| R-Pulsar [85, 87] | ✓ | ✓ | ✓ | ✓ | ✓ | ✓ |
| Fogflow [44] | | ✓ | ✓ | ✓ | | |
| SiteWhere [8] | | ✓ | | ✓ | | |
| EdgeX [28] | | ✓ | | ✓ | | |
| PiCasso [66] | | | ✓ | ✓ | | |
| Hua-Jun Hong et. al. [56] | | | | ✓ | | |
| Cloud4IoT [81] | ✓ | | ✓ | | | |
| CEFIoT [59] | | | | ✓ | | |
| Foggy [101] | | | | ✓ | | |
| ISYMPHONY [100] | | | | ✓ | | |

Table 8. (Appendix 1) Data processing layer and design goals for all the commercial and academic edge middleware available.

| System | Data Ingestion | | | Data Analysis | | | Data Storage | | | Data Query | | |
|---|---|---|---|---|---|---|---|---|---|---|---|---|
| | Real-Time | Distributed | Scalable | Real-Time | Distributed | Scalable | Real-Time | Distributed | Scalable | Real-Time | Distributed | Scalable |
| AWS Greengrass [33] | | ✓ | ✓ | ✓ | ✓ | ✓ | | | | | | |
| Azure IoT Edge [75] | | ✓ | ✓ | ✓ | ✓ | ✓ | | ✓ | ✓ | | ✓ | ✓ |
| EAaaS [99] | ✓ | ✓ | ✓ | ✓ | ✓ | ✓ | | | | | | |
| Google Cloud IoT [20] | | ✓ | ✓ | | ✓ | ✓ | | | | | | |
| Cisco IoT Cloud Connect [12] | | ✓ | ✓ | ✓ | ✓ | ✓ | | | | | | |
| Everyware IoT [14] | | ✓ | ✓ | | ✓ | ✓ | | | | | | |
| Predix [48] | | ✓ | ✓ | | ✓ | ✓ | | | | | | |
| Bosch IoT [22] | | ✓ | ✓ | | ✓ | ✓ | | | | | | |
| Yanzi et. al. [5] | | ✓ | ✓ | | ✓ | ✓ | | | | | | |
| R-Pulsar [85, 87] | ✓ | ✓ | ✓ | | | | ✓ | ✓ | ✓ | ✓ | ✓ | ✓ |
| FogHorn [15] | | ✓ | ✓ | | ✓ | ✓ | | | | | | |
| GeeLytics [43] | | ✓ | ✓ | ✓ | ✓ | ✓ | ✓ | ✓ | ✓ | | ✓ | ✓ |
| Fogflow [44] | | ✓ | ✓ | ✓ | ✓ | ✓ | ✓ | ✓ | ✓ | | ✓ | ✓ |
| OpenMTC [13] | | ✓ | ✓ | ✓ | ✓ | ✓ | ✓ | ✓ | ✓ | | ✓ | ✓ |
| SiteWhere [8] | | ✓ | ✓ | | ✓ | ✓ | | | | | | |
| SmartThings [11] | | ✓ | ✓ | | ✓ | ✓ | | | | | | |
| Kaa [16] | ✓ | ✓ | ✓ | ✓ | ✓ | ✓ | ✓ | ✓ | ✓ | ✓ | ✓ | ✓ |
| Samsung Artik | | ✓ | ✓ | | ✓ | ✓ | | | | | | |
| Ayla Network [9] | | ✓ | ✓ | | ✓ | ✓ | | | | | | |
| Altair SmartWorks [18] | | ✓ | ✓ | | ✓ | ✓ | | | | | | |
| EdgeX [28] | | ✓ | ✓ | | ✓ | ✓ | | | | | | |
| PiCasso [66] | | | | | ✓ | ✓ | | | | | | |
| Hua-Jun Hong et. al. [56] | | ✓ | ✓ | | ✓ | ✓ | | ✓ | ✓ | | ✓ | ✓ |
| Cloud4IoT [81] | | ✓ | ✓ | | ✓ | ✓ | | ✓ | ✓ | | | |
| Nebulae [25] | | ✓ | ✓ | | ✓ | ✓ | | | | | | |
| FogGIS [37] | | ✓ | ✓ | | ✓ | ✓ | | ✓ | ✓ | | ✓ | ✓ |
| FOG-engine [74] | | ✓ | ✓ | | ✓ | ✓ | | | | | | |
| CEFIoT [59] | | ✓ | ✓ | | ✓ | ✓ | | | | | | |
| SAVI-IoT [61] | | ✓ | ✓ | | ✓ | ✓ | | | | | | |
| Foggy [101] | | ✓ | ✓ | | ✓ | ✓ | | | | | | |
| ISYMPHONY [100] | | ✓ | ✓ | | ✓ | ✓ | | | | | | |
| Macchina.io [17] | | ✓ | ✓ | | ✓ | ✓ | | ✓ | ✓ | | ✓ | ✓ |
| Clearblade [3] | | ✓ | ✓ | | ✓ | ✓ | | | | | | |
| IBM Watson IoT [24] | | ✓ | ✓ | ✓ | ✓ | ✓ | | | | | | |

Table 9. (Appendix 1) Service layer and design goals for all the commercial and academic edge middleware available.

| System | Rule Engine | | Programming Model | | Workflow Orchestrator | | |
|---|---|---|---|---|---|---|---|
| | Scalable | Low Overhead | Expressive | Extensible | Dynamic | Scalable | Low Overhead |
| AWS Greengrass [33] | ✓ | ✓ | ✓ | ✓ | | | |
| Azure IoT Edge [75] | ✓ | ✓ | ✓ | ✓ | | | |
| EAaaS [99] | ✓ | ✓ | ✓ | ✓ | | | |
| Google Cloud IoT [20] | ✓ | ✓ | ✓ | ✓ | | | |
| Cisco IoT Cloud Connect [12] | | | ✓ | ✓ | | | |
| Everyware IoT [14] | ✓ | ✓ | ✓ | ✓ | | | |
| Predix [48] | ✓ | ✓ | ✓ | ✓ | | | |
| Bosch IoT [22] | ✓ | ✓ | ✓ | ✓ | | | |
| Yanzi et. al. [5] | | | ✓ | ✓ | | | |
| R-Pulsar [85, 87] | ✓ | ✓ | ✓ | ✓ | ✓ | ✓ | ✓ |
| FogHorn [15] | ✓ | ✓ | ✓ | ✓ | | | |
| GeeLytics [43] | | | ✓ | ✓ | | | |
| Fogflow [44] | ✓ | ✓ | ✓ | ✓ | ✓ | ✓ | ✓ |
| OpenMTC [13] | | | ✓ | ✓ | | | |
| SiteWhere [8] | ✓ | ✓ | ✓ | ✓ | | | |
| SmartThings [11] | ✓ | ✓ | ✓ | ✓ | | | |
| Kaa [16] | ✓ | ✓ | ✓ | ✓ | | | |
| Samsung Artik [26] | | | ✓ | ✓ | | | |
| Ayla Network [9] | ✓ | ✓ | ✓ | ✓ | | | |
| Altair SmartWorks [18] | | | ✓ | ✓ | | | |
| EdgeX [28] | ✓ | ✓ | ✓ | ✓ | | | |
| PiCasso [66] | | | ✓ | ✓ | ✓ | ✓ | ✓ |
| Hua-Jun Hong et. al. [56] | | | ✓ | ✓ | | | |
| Cloud4IoT [81] | | | ✓ | ✓ | ✓ | ✓ | ✓ |
| Nebulae [25] | ✓ | ✓ | ✓ | ✓ | | | |
| FogGIS [37] | | | ✓ | ✓ | | | |
| FOG-engine [74] | ✓ | ✓ | ✓ | ✓ | ✓ | ✓ | ✓ |
| CEFIoT [59] | | | ✓ | ✓ | | ✓ | |
| SAVI-IoT [61] | | | ✓ | ✓ | | ✓ | |
| Foggy [101] | | | ✓ | ✓ | ✓ | | ✓ |
| ISYMPHONY [100] | | | ✓ | ✓ | ✓ | ✓ | ✓ |
| Macchina.io [17] | | | ✓ | ✓ | | | |
| Clearblade [3] | ✓ | ✓ | ✓ | ✓ | | | |
| IBM Watson IoT [24] | ✓ | ✓ | ✓ | ✓ | | | |